\documentclass[preprint,5p,twocolumn,authoryear,times]{elsarticle}

\usepackage[labelfont=bf,singlelinecheck=false]{caption}
\usepackage[font=footnotesize,labelfont=bf]{subcaption}
\usepackage{amsfonts}
\usepackage[cmex10]{amsmath}
\usepackage{graphicx}
\usepackage{diagbox}
\usepackage{footmisc}
\usepackage{makecell}
\usepackage{multirow}
\usepackage{booktabs}
\usepackage[flushleft]{threeparttable}
\usepackage{url}
\usepackage{stfloats}
\usepackage{epsfig}
\usepackage{siunitx}
\usepackage{flushend}

\journal{Elsevier}

\biboptions{authoryear}

\usepackage[x11names]{xcolor}
\usepackage[colorlinks]{hyperref}
\usepackage[nameinlink,noabbrev]{cleveref}
\AtBeginDocument{\hypersetup{citecolor=DodgerBlue4,allcolors=DodgerBlue4}}

\begin{document}

\begin{frontmatter}

\title{$Laoco\ddot{o}n$: Scalable and Portable Receipt-free\\ E-voting Protocol without Untappable Channels}

\author[1]{Shufan Zhang}
\author[1]{Hu Xiong\corref{mycorrespondingauthor}}


\address[1]{School of Information and Software Engineering,University of Electronic Science and Technology of China, Chengdu, China}

\cortext[mycorrespondingauthor]{Corresponding author}
\ead{xionghu.uestc@gmail.com}


\begin{abstract}
Vote-buying and voter-coercion are the impending threats when deploying remote online voting into large scale elections. With a policy of carrot and stick, it will encourage voters to deviate from honest voting strategy and spoil the democratic election. To deal with this problem, many voting protocols proposed their solutions with the notion of receipt-freeness. However, existing receipt-free voting protocols either rely on some impractical assumptions as untappable communication channel, or are burden with heavy voter-side computation and quadratic tallying complexity.
In this paper, we present $Laoco\ddot{o}n$, a brand new cryptographic voting protocol which is practical and light-weight to be deployed in large scale online elections. By taking advantage of proxy re-encryption, our protocol can defend vote-buying attacks. Furthermore, we introduce a new property, candidate-adaptiveness, in electronic voting which refers to as every candidate knows the real-time vote number towards himself, while he knows nothing about others, nor he buys votes. We prove the correctness of our protocol and evaluate the performance with experimental results. Finally we advance some open problems which will be coped in our future work.
\end{abstract}


\end{frontmatter}


\section{Introduction}

At first glance, remote electronic voting (REV) allows voters to vote with no spacial restriction, and encourages greater voter turnout. Furthermore, electronic means can help to reduce the human cost and accelerate both the computation of election result and the democratic process.
A closer look \citep{gibson2016review}, however, suggests that remote electronic voting still fails to gain world wide acceptance. In most countries or regions, only optical ballot scanners or ATM-like voting machines (i.e. Direct Recording Electronic machines) are adopted as components in government elections.

One point is that \citep{rubin2002security}, though information technology makes voting more efficient and convenient, the threats of \emph{vote-buying} and \emph{voter-coercion} are, in the meantime, exacerbated.
In elections when candidates are in a statistical tie, unaffiliated voters will become the battlefield. Then vote-buying is a disreputable but efficient approach for the candidates. What makes matter worse, voter may also have the willing to sell his/her ballot for a higher price. \footnote{One example is the notorious forum for ballot auctioning: \url{ www.vote-auction.net}.} In elections of poor nations, unjust voting campaign may even evolve into political coercion which endangers safety of voters. These attacks exert negative influence on fairness of voting and the aftermath is catastrophic to democratization. Paul Collier called this "democrazy" \citep{collier2011wars} and delineated pervasive political coercion and vote-buying phenomena in African elections.

Facilitated by anonymous payment means and cryptographic tools, modern adversaries (malicious candidates) can bribe or bludgeon voters efficiently and secretly in a distance. Since the attacks are easier to carry out but harder to trace back, the defence should be taken into consideration when a remote electronic voting system is designed. The corresponding terminology in electronic voting literatures is \emph{privacy}, which leads to several evaluation criteria. One basic binary classification results in \emph{ballot secrecy} and \emph{receipt-freeness}. The former one means that a voter \emph{can} keep his/her choice private so that no "voter-vote" relationship can be externally observed, while the latter requires that voter \emph{must} keep his/her choice private \citep{hirt2000efficient}. That is, the voter is incapable to prove to any third-party the way he voted, and therefore, the vote-buying strategy will be abandoned. Receipt-freeness brought a intriguing crux in remote electronic voting study, and has always been a research hot-spot. Many investigations 
have discussed about this topic and achieved it with different assumptions and design philosophy. Nevertheless, how to achieve receipt-freeness efficiently and practically, is still a challenging task.

\subsection{Related Work}
\label{reviews}



The notion of \emph{receipt-freeness} was first introduced in 1994 by Benaloh and Tuinstra \citep{benaloh1994receipt}. Their scheme, based on the assumption of a physical voting booth and homomorphic encryption, was proved neither convenient nor receipt-freeness as originally envisaged \citep{hirt2000efficient}. Since then, researchers commenced the three-decade long study on receipt-freeness. One stream of research leads to incoercible multi-party computation (MPC), e.g. \citep{canetti1996incoercible,hao2010anonymous}. Though they provide perfect receipt-freeness, as the off side of coin, they are naturally limited to small-scale (a.k.a. boardroom) voting. In the context of large-scale receipt-free voting, researchers attempted various privacy-preserving primitives, such as homomorphic encryption, mix-net, blind signature and zero knowledge proof, just to name a few. After early attempts, a consensus has been reached: receipt-freeness does not stem directly from the primitives, but from some tricky usages or combinations with additional assumptions.

Mix-net is one commonly proposed tool to backup secure electronic voting protocols \citep{chaum1981untraceable,fujioka1992practical,sako1995receipt,okamoto1997receipt,ohkubo1999improvement,boneh2002almost,chaum2005practical,ryan2006pret,ryan2009pret,carroll2009secure,wu2014electronic}. Roughly speaking, mix-net is a set of servers which takes a collection of ciphertexts as input, shuffles and outputs messages that are unlinkable to the incoming ones. Some variants of mix-net, e.g. the onion routing (TOR) \citep{dingledine2004tor}, are used to achieve an anonymous communication channel between voter and voting authorities \citep{fujioka1992practical,juels2005coercion}. Besides, mix-net also serves as backbone in some voting protocols \citep{chaum1981untraceable,sako1995receipt,boneh2002almost,juels2005coercion,chaum2005practical,ryan2006pret}. In a mix-net based voting protocol, voter anonymity can be easily achieved. For example, ballots can be treated as ciphertext and cast to the bulletin board (i.e. a public-accessed database). After the voting process, mix servers collect the ballots as input, decrypt with shuffling and output plaintext messages of voter choices. Provided at least one server is trustworthy, any adversary will not tell which vote was cast by which voter. It is regrettable that the example is not receipt-free. Simply by furnishing the voter with a pre-determined ciphertext, the adversary can verify whether the voter cast the particular vote.

Protocols based on homomorphic encryption \citep{benaloh1994receipt,sako1995receipt,hirt2000efficient,chow2008robust,wen2009masked,yi2013practical,xia2018framework} share the same vulnerability with those based on mix-net in achieving receipt-freeness. Different from mix-net based protocols which are more suitable for elections with multiple candidates, homomorphic encryption based protocols are recommended to be deployed in "YES/NO" voting due to its linear growth cost related to the number of candidates \citep{aditya2003secure}. Homomorphic encryption is utilized to facilitate tallying process. Thanks to the addictive or multiplicative \footnote{As for the comparison between addictive and multiplicative homomorphic encryption used in voting protocols, interested reader can refer to \citep{peng2004multiplicative}.} homomorphic feature, ballots can be tallied while no single one is recovered. Hence voter anonymity is preserved.

Generally speaking, receipt-freeness achieved in mix-net or homomorphic encryption based protocols comes with some assumptions. One example is the \emph{physical voting booth} \citep{benaloh1994receipt,yu2018platform}, where voters has perfect privacy in communication with voting authorities. Following investigations take \emph{untappable channel}, a dedicated communication channel that is perfectly secret and free from eavesdropping, as a replacement to physical voting booth. Stated roughly, physical voting booth is a full-duplex untappable channel, while untappable channel can be unidirectional from voter to voting authorities \citep{okamoto1997receipt} and vice versa \citep{sako1995receipt,hirt2000efficient}. With these assumptions, voting authorities can add some voter-unknown randomness into the ballot before it is published to the bulletin board. Since this process is out of observation from adversaries, voter can not convince adversaries that he vote as instructed. Thus, receipt-freeness is achieved.

Though similarity exists in protocols based on mix-net and homomorphic encryption, Hirt and Sako \citep{hirt2000efficient} believe that homomorphic encryption works more efficient than mix-net in receipt-free voting protocols since mix-net requires a heavier processing load for tallying. Their protocol is followed by a more efficient one \citep{baudron2001practical} and these two are still of the most efficient voting protocols to date. Among those receipt-free protocols with homomorphic encryption, Masked Ballot \citep{wen2009masked} distinguishes itself by a tricky approach of splitting ballot. In this protocol, the digital representation of a candidate is split into two parts. One, which is called the mask, is cast in registration stage and another is cast in voting stage. Both ballots are encrypted and the tallying correctness is guaranteed by the homomorphic feature of encryption. By using the mask, vote-buyers are blocked from the voting process. However, a recent study \citep{xia2018framework} showed that Masked Ballot is vulnerable to shifting vote attack and proposed an improvement based on it.

Blind signature well protects anonymity of voters. Okamoto \citep{okamoto1997receipt} proposed a receipt-free protocol, based on blind signature, which repairs yet another voting protocol that the author himself proposed earlier \citep{okamoto1996electronic}. Blind signature based protocol relies on an \emph{anonymous untappable channel}, which is an even stronger assumption. Furthermore, in registration phase, voter needs more than one-round interaction with voting authorities via anonymous untappable channel, which make the protocol extremely impractical in real world. Hence few recent study constructs receipt-free protocol with blind signature.

As demonstrated in \citep{lee2002receipt}, the assumption of untappable channels is not only impractical, but also unsuitable for voting over Internet. \footnote{The Masked Ballot protocol \citep{wen2009masked} with the assumption of one-way untappable channel claims to design for \emph{online voting}. In fact, their protocol is divided into two parts where only the voting part can be put online. As the authors themselves pointed, the registration part in their protocol, which uses the untappable channel, can not be implemented over the Internet.}
Nowadays, online voting and mobile voting are in trend \citep{jan2001design,li2008electronic}. Cryptographers attempt to seek for more practical substitutes.
Inspired by \citep{magkos2001receipt}, Lee and Kim \citep{lee2002receipt} proposed to use tamper-resistant hardware to replace impractical assumptions in real world. However, trusted hardware may cause other security concerns and is too costly to be deployed in large scale elections. In 2005, Juels, Catalano and Jakobsson \citep{juels2005coercion} suggested that the minimum requirement for receipt-freeness is an anonymous channel. Their proposal has been the most practical receipt-free protocol hitherto and was implemented to the real world later \citep{clarkson2008civitas}. Even so, there are still defects on its quadratic complexity in tallying procedure \citep{araujo2010practical,spycher2011new} and the excessively heavy voter-side computation \citep{xia2018framework}.

Receipt-freeness is not free. As noted in \citep{wang2017review}, receipt-freeness is somewhat conflict with \emph{verifiability}, another essential property which means the system allows individual voters to confirm, by a verifiable receipt, that their votes are correctly decoded and tallied.
Verifiability is to create a verifiable receipt to voters, while receipt-freeness thwarts a preference provable receipt.
Some electronic voting systems \citep{adida2008helios,adida2009electing,bulens2011running}, including a recently proposed open-source implementation \citep{haenni2017chvote}, achieve verifiability at the cost of sacrificing receipt-freeness.
A formal proof \citep{chevallier2010some}, shows that an electronic voting system cannot simultaneously achieve verifiability and receipt-freeness, unless some strong assumptions, such as untappable channels, are available. As the consequence, the proposal of Juels et al. \citep{juels2005coercion} and other practical protocols (such as \citep{araujo2008practical}), cease to be universal verifiable.

\subsection{Our Contribution}
From the above analysis, we can observe the state-of-the-art approaches to achieve receipt-freeness in a voting protocol. In this paper, we propose $Laoco\ddot{o}n$ \footnote{$Laoco\ddot{o}n$ is derived from the name of a Trojan priest, who said the famous line "Beware of Greeks bearing gifts" in Virgil's poem. This name reflects our main goal: Beware of the malicious \emph{Vote-Buyers} bearing gifts. }, the first proxy re-encryption based e-voting protocol which enjoys multiple features. The novel protocol meets the attractive features as follows:
\begin{itemize}
  \item[1.] Our voting protocol is \emph{receipt-free} and \emph{practical}. Different from former investigations, in our study, receipt-freeness is achieved by a brand new approach which no longer relies on any impractical physical assumptions. It only takes an anonymous channel as necessity, which is demonstrated in \citep{juels2005coercion} as the minimal requirement for achieving receipt-freeness. The details of security analysis of receipt-freeness and other properties are also discussed.

  \item[2.] Our voting protocol follows a new property called \emph{candidate-adaptiveness}. We consider it reasonable that during the election process candidate can adopt some campaign strategies such as addressing some speeches. To achieve this property, candidate is permitted to know the real time portion of ballots towards him. Since our scheme is meanwhile receipt-freeness, voters' privacy will not leak out and vote selling will not occur when candidate-adaptiveness is brought in. Furthermore, our protocol is \emph{more suitable for elections with a large number of candidates}, and is \emph{suitable for both 1-out-of-L and k-out-of-L candidates elections}.
  \item[3.] Our voting protocol is \emph{scalable} and \emph{can be deployed in online voting or mobile voting}. As our evaluation in the experimental result, the voter-side calculation is light-weight which can encourage higher voter turnout to make a election scalable. Furthermore, in our proposal there is no restricted physical location where voter cast his ballot. Hence with the receipt-freeness property, our protocol is suitable for online voting or mobile voting.

\end{itemize}

\subsection{Organization}

The remainder of this paper is organized as follows. The cryptographic tools, proxy re-encryption and multi-designated verifiers signature, are discussed in \Cref{sec_crypto_tools}. The model of electronic voting protocol is introduced in \Cref{sec_voting_model}. In \Cref{sec_proposal}, we delineate the high level rationale and technique details of our proposed protocol, $Laoco\ddot{o}n$. We give an analysis of $Laoco\ddot{o}n$ and a functionality comparison with other receipt-free protocols in \Cref{sec_analysis}. Finally, we conclude our work and advance some open problems in \Cref{sec_conclusion}.

\section{Background}
\label{sec_crypto_tools}



\subsection{Proxy Re-Encryption}
Proxy re-encryption (PRE) was first introduced by Blaze, Bleumer and Strauss \citep{blaze1998divertible}. As an extension of public key encryption, proxy re-encryption enables a honest-but-curious proxy to transform a ciphertext encrypted under delegator's public key into another ciphertext under delegatee's public key without leaking any information of corresponding message. To perform the transformation correctly, proxy obtains as little information as necessary, where the specific information is called a re-encryption key (a.k.a transformation key). To classify the numerous PRE schemes, Blaze et al. provide two methods. One classification results from the allowed times of transformation. If a PRE scheme allows proxy to repeatedly transform a ciphertext, e.g. from Alice to Bob, then from Bob to Charlie, it is multi-use; otherwise, it is single-use. The other classification results from the allowed direction of transformation. If a PRE scheme allows proxy to use the same re-encryption key to transform ciphertext from Alice to Bob, and vice versa, it is bidirectional; otherwise, it is unidirectional.

PRE can be used in such various application scenarios as email forwarding \citep{blaze1998divertible}, distributed file systems \citep{ateniese2006improved}, digital rights management \citep{taban2006towards} and cloud computing \citep{yu2010achieving} to avoid heavy workload of the decryption-then-encryption approach. To the best of our knowledge, no previous investigation regards electronic voting as a potential scenario for PRE schemes. Fundamentally different from the previous applications of PRE, there exists no decryption-then-encryption problem in voting protocols. However, by applying a special PRE scheme in a special approach, one essential property for voting, receipt-freeness, is achieved practically and with light-weight user-side calculation.

\subsubsection{Key-private Proxy Re-encryption}

In 2009, Ateniese et al. \citep{ateniese2009key} proposed the concept of key-private (or anonymous) proxy re-encryption. This additional useful property of proxy re-encryption refers to that adversary (proxy) can not deduce the identity of both delegator and delegatee from the re-encryption key. That is, information as "who was speaking privately with whom" \citep{ateniese2009key} or "who is using the re-encryption service" \citep{shao2012anonymous} can not be extracted by adversary when the proxy is compromised.

In practice, the special PRE scheme we applied is the key-private proxy re-encryption from Ateniese et al. \citep{ateniese2009key}. Their proposed scheme is a single-use unidirectional key-private PRE scheme (SU-KP-PRE) with CPA security. Despite that CPA security property is much weaker than CCA security, it is enough for an electronic voting protocol. That is because, in a round of election, voters and voting authorities need not to answer the decryption queries.

\subsubsection{Function Notions}
 A single-use unidirectional key-private PRE scheme (SU-KP-PRE) contains a set of P.P.T. (probably polynomial time) algorithms (Setup, KeyGen, ReKeyGen, Enc, ReEnc, Dec):
 \begin{description}
   \item[\textbf{Setup:} ] $\mathsf{par} \gets \textbf{Setup}(1^k)$ takes an input of security parameter $k$, returns global public parameter $\mathsf{par}$.

   \item[\textbf{Key Generation:} ] $(PK_i, SK_i) \gets \textbf{KeyGen}(\mathsf{par})$ takes an input of global public parameter $\mathsf{par}$, returns a pair of public/secret key.

   \item[\textbf{Re-Key Generation:} ] $rk_{i \to j} \gets \textbf{ReKeyGen}(SK_i, PK_j)$ takes an input of the secret key of $i$, and the public key of $j$ where $j\neq i$, returns a re-encryption key from i to j.

   \item[\textbf{Encryption:} ] $c_i \gets \textbf{Enc}_{PK_i}(m)$ takes an input of the public key of $i$, and a message $m$, returns a ciphertext $c$.

   \item[\textbf{Re-Encryption:} ] $c_j \gets \textbf{ReEnc}_{rk_{i \to j}}(c_i)$ takes an input of ciphertext under public key of $i$, and a re-encryption key from $i$ to $j$, returns ciphertext under public key of $j$.

   \item[\textbf{Decryption:} ] $m \gets \textbf{Dec}_{SK_i}(c_i)$ takes an input of the secret key of $i$, and a ciphertext under the public key of $i$, returns the original plaintext message $m$.
 \end{description}

Since the security of our voting protocol is related to the construction of PRE scheme, we present the algorithm details of Ateniese et al.'s \citep{ateniese2009key} in \ref{APP_PRE_Details}.

\subsection{Multi-Designated Verifiers Signature}

As a generalization of Designated Verifier Signature (DVS) introduced by Jakobsson et al. \citep{jakobsson1996designated}, the notion of Multi-Designated Verifiers Signatures (MDVS) was discussed in the rump session of Crypto'03 and later formalized in \citep{laguillaumie2004multi}. MDVS allows a signer to issue a signature whose validity can only be verified by a specific set of verifiers chosen by the signer. DVS/MDVS are useful in voting protocols \citep{juels2005coercion}, since the designated verifier(s) can forge a indistinguishable fake signature to defraud adversaries. MDVS provides stronger anonymity than DVS \citep{laguillaumie2007multi}, as the number of designated verifiers extended from one to many.

\subsubsection{Function Notions}

A MDVS scheme contains a set of P.P.T. algorithms (Setup, KeyGen, Sign, and Verify). Since the algorithm of Setup and KeyGen is somehow redundant to the algorithms in PRE scheme, we only describe the Sign and Verify functions.

\begin{description}
  \item[\textbf{Signature:} ] $s \gets \textbf{Sign}_{SK_S, \mathcal{V}}(m)$ takes an input message $m$, a secret key $SK_S$ of signer $S$, and a set of designated verifiers $\mathcal{V}$, and returns a signature $s$, which is a DVS of $m$ signed by $S$.
  \item[\textbf{Verification:} ] $1/0 \gets \textbf{Verify}_{PK_S, SK_V, \mathcal{V}}(s, m)$ takes an input of message/signature pair $(s, m)$, the public key $PK_S$ of signer $S$, the secret key of verifier $SK_V$ whose identity is involved in $\mathcal{V}$, and outputs $1/0$ which represented whether the signature is valid or invalid.
\end{description}

\section{Model of Electronic Voting}
\label{sec_voting_model}

\subsection{Entities}

As illustrated in \Cref{fig_sequence}, five entities are involved in $Laoco\ddot{o}n$. Respectively, they are the administrator, the voters, the proxy, the bulletin board and the candidates. The detailed definitions of the entities are showed as follows:

\textbf{Administrator:} As the election holder, administrator is to organize or control the voting process by initializing the system parameters and triggering different phases of an election. We use $(y_A, x_A)$ to denote the pair of public/secret key of administrator.

\textbf{Voters:} A set of legal voters, which are denoted by $\vec{L}_v = \{V_1, V_2, \dots, V_n\}$, are authorized to vote for their preferred candidates with independent judgements. At any time in the election, legal voters \emph{can} be corrupted by adversaries, and turn to malicious ones which are willing to sell ballots or vote more than once. We assume that all voters hold a long-term public/secret key pair $(y_V, x_V)$ that represents their identities.

\textbf{Proxy:} As a role played by independent authorities, proxy is a \emph{honest-but-curious} entity that processes on re-encryptable ciphertexts and handles the ballots. As shown in \Cref{fig_sequence}, proxy serves not only in registering (Step 5), but also in ballot processing (Step 11).

\textbf{Candidates:} A set of candidates, which are denoted by $\vec{L}_c = \{C_1, C_2, \dots, C_t\}$, campaign to obtain more votes and compete to win the election. Unlike most voting protocols, candidates, in our proposal, participate in tallying process where they uncover the encrypted transactions and prove the number of votes that themselves obtain.

\textbf{Bulletin board:} Denoted as $\mathcal{BB}$, bulletin board is a \emph{tamper-resistant} and \emph{append-only} database \citep{heather2008append} which can be publicly accessed. Commonly, the existence of $\mathcal{BB}$ is to simulate a broadcast communication channel and to model a public memory for achieving verifiability.

Since no untappable channel is existed in $Laoco\ddot{o}n$, which means the threat of eavesdropping, there is theoretically no need to have end-to-end communication. Hence all voting related information can be uploaded to $\mathcal{BB}$. Note that we also maintain an end-to-end channel for Step 4$\sim$6, only for reducing the complexity.

\subsection{Functions}

\begin{figure}[!t]
    \centering
    \includegraphics[width=3.6in]{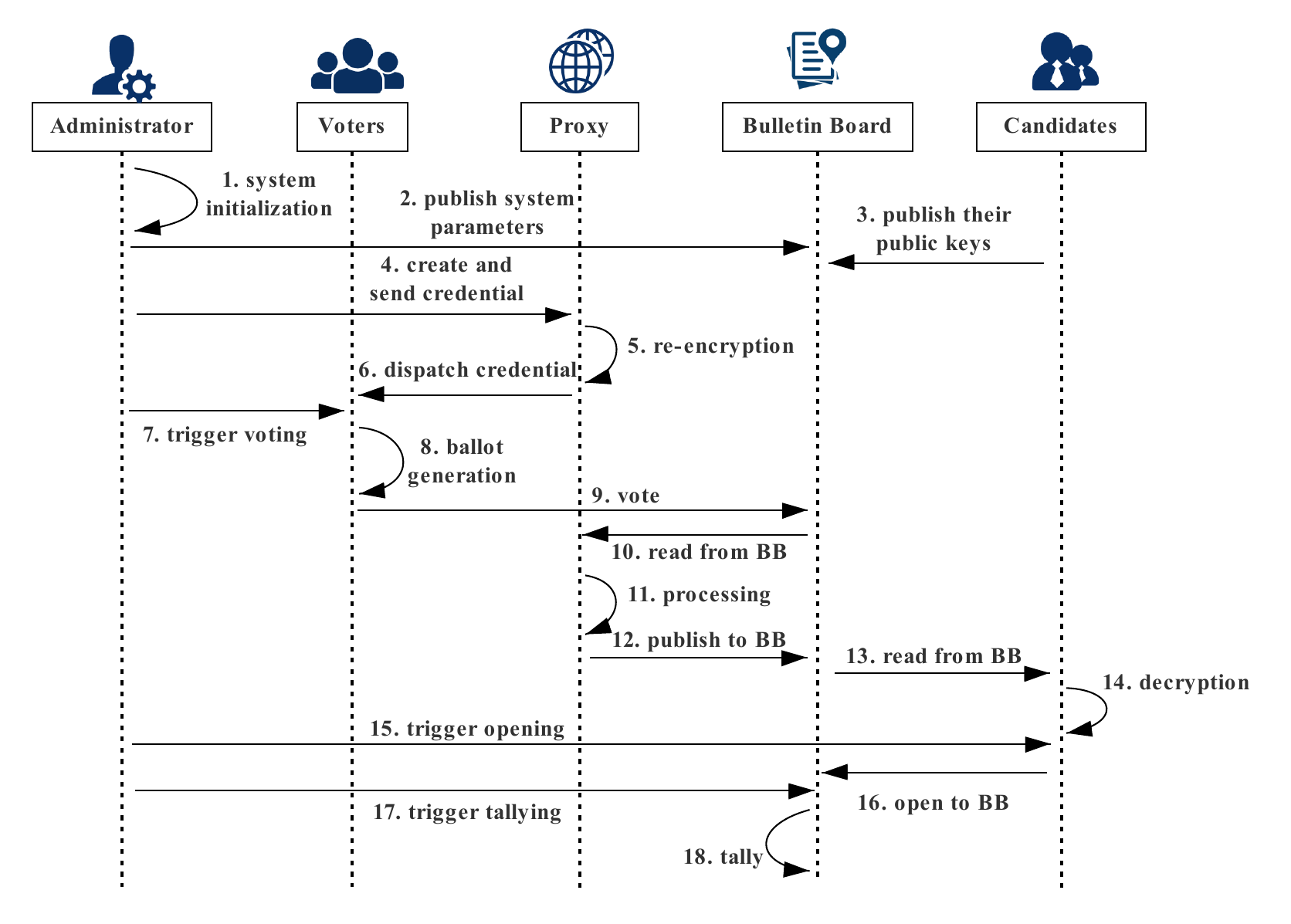}
    \caption{System sequence diagram.}
    \label{fig_sequence}
\end{figure}

We assume that $\vec{L}_v$ is a list of all legal voters, $\vec{L}_k$ is the corresponding list of re-encryption keys used in Step 5, and there exist n (n $\textgreater$ 2) candidates in the election. The communication process among the entities is as shown in \Cref{fig_sequence} which comprises 18 steps. We divide the whole process into the following five phases.


\textbf{System Setup.} In this phase, voting administrator runs the Setup algorithms of used cryptographic primitives, and publishes the generated public parameters to the bulletin board. Administrator also uploads his self-signed public key certificate and the list of $\vec{L}_k$ used for credential dispatching. Then all candidates publish their public key for this round of election.

\textbf{Credential Dispatching.} In this phase, anonymous voting credentials are created and issued by voting administrator, and dispatched by proxy after a secret shuffle. With the help of proxy re-encryption, no voting authorities will learn the relationship between a particular voter and the plaintext credential.

\textbf{Ballot Casting.} Voter generates a ballot for the favoured candidate. The ballot is later casted to the bulletin board via anonymous channel. For any ballot that appears on $\mathcal{BB}$, proxy chooses some random message and processes on the ballot. The result, namely the voting transaction, which is a ciphertext under candidate's public key, is also published to $\mathcal{BB}$.

\textbf{Ballot Opening and Tallying.} Candidates engage to open the ballots towards them. By monitoring bulletin board, candidate roughly learns how it positioned versus its competitors. As \emph{Tallying Date} approaches, voting administrator informs all candidates publish their secret key for this round of election. A pre-designed tallying script on bulletin board will be triggered on that date. The final tallying result will be published on the bulletin board then.

\textbf{Ballot Verifying and Auditing.} This is a optional phase to the whole election. In a period of time after publishing the tallying result, transcript of the bulletin board requests for open auditing. In the \emph{verifiability achieved version} of $Laoco\ddot{o}n$, voter can claim to voting administrator if it finds that its ballot is not tallied to the result.

\subsection{Design Goals}
  Compared with the existing works, our novel voting protocol, $Laoco\ddot{o}n$, should satisfy the following properties:


 \textbf{Completeness.} The counting of the valid ballots is accurate when the protocol is followed by all participants.

 \textbf{Ballot secrecy.} Voters' anonymity will be guaranteed by this protocol, i.e. ballots will not leak information about identities of both voter and the preferred candidate even if proxy is corrupted.

 \textbf{Unreusability.} Ballot from the same legitimate voter will not be counted twice.

 \textbf{Eligibility.} Only legitimate voters' ballots will be counted.


 \textbf{Vote-and-go.} A voter can go off-line once his ballot is casted. That is, voter will not engage in opening ballots.


 \textbf{Efficiency.} The computation and communication consumption of the voting scheme is not too huge to allow voters vote on tablets or mobile phones.

 \textbf{Mobility.} There is no restriction on the designated location in which voter can cast its ballot.

 \textbf{Receipt-freeness.} A voter cannot prove to someone that she voted in a certain approach.


 \textbf{Verifiability.} Any individual voter can confirm that its choice has been correctly encoded and tallied. \footnote{\emph{Verifiability} is served as an extended property in our proposed protocol.}

 In $Laoco\ddot{o}n$, we weaken the fairness property that is mentioned as additional requirement in some voting schemes to \emph{Voter-fairness}. Fairness property is generally defined as that no partial result can be computed by \emph{anyone} before the end of election. We notice that in some circumstances it is available for candidate to know how many ballots are towards him in the election process and to adaptively implement some campaign strategies. Hence we weaken the fairness property and propose a new voting requirement to satisfy this environment called \emph{Candidate-adaptiveness}.

 \textbf{Voter-fairness.} No voter can break the protocol to get any results before ballots opening and tallying.

 \textbf{Candidate-adaptiveness.} Each candidate in election can only know how many ballots are toward it in the election process but get no knowledge about the exact number of ballots towards other explicit competitor. This property is useful when the number of candidates in an election is more than two.

\subsection{Encoding of Ballots}

In our proposed protocol, ballots have two different forms. One is what voter generates and casts to the proxy, and another is what proxy publishes to the bulletin board.

The former one, called ballot, is the re-encryption key in proxy re-encryption scheme. It is generated by putting voter's voting pseudonym secret key and preferred candidate's public key into the re-encryption key generation algorithm \textbf{ReKeyGen}. If a voter with pseudonym $i$ favours candidate $C_x$, the ballot it casts can be denoted as $Ballot \gets rk_{i \to C_x}$, where $rk_{i \to C_x} := ReKeyGen(SK_i, y_{C_x})$.

The latter one, named voting transaction, is the re-encrypted ciphertext under the public key of candidate. The corresponding plaintext is the current timestamp selected by proxy. It is denoted as $\delta_j^\prime \gets ReEnc_{Ballot}(\delta_j)$, where $\delta_j$ is the encrypted timestamp $\delta_j \gets Enc_{PK_i}(Stp)$.

\section{The Proposed Protocol: $Laoco\ddot{o}n$}
\label{sec_proposal}

\subsection{High Level Idea}

Based on the key-private proxy re-encryption scheme and other cryptographic tools, we now present the high-level design rationale of our proposed protocol, $Laoco\ddot{o}n$.

To block misbehaved or illegal voters without compromising their anonymity, which means one essential goal, \emph{unreusability}, some cryptographic means that provide anonymous authentication \citep{tsang2008perea} should be kept in voting protocols.
Our protocol inherits the thought of anonymous credential from some previous investigations \citep{juels2005coercion,tsang2007blacklistable,tsang2011nymble,mateu2014constructing}, but works slightly different.
In our protocol, a legal voter, say Alice, holds two pairs of public/secret keys, i.e. a pair of long-term keys that represents her identity and a pair of short-term keys that is only used in a temporal round of voting process. The pair of short-term key, also known as Alice's voting pseudonym, is dispatched with an voting credential by voting authorities before each round of voting process. Generally speaking, Alice retain anonymity by the approach of receiving pseudonym with her long-term keys and voting with her short-term keys.

Anonymity is brought by pseudonym, while authentication results in voting credential. The voting credential is comprised of a short-term public key with a digital certificate from voting administrator. The usage of anonymous credential shifts authentication from voting process to an earlier phase. In the ballot casting phase, a ballot is received as a legal one so long as it is cast with an unused legal credential. In the earlier phase, voting authority issues credentials and dispatches them to legitimate voters. If one voting authority is in charge of both issuing and dispatching, it will readily know which pseudonym has been distributed to which voter. Hence in our design, we separate the responsibility into two parts. Voting administrator works as the issuer and proxy works as dispatcher. A PRE scheme is used in this process to encrypt the credential and thwart curious proxies. Unless no corruption of all authorities happens, the crisis of confidence is eliminated. It is noteworthy that no untappable channel is needed here. Since the plaintext credential is protected by PRE, no one but the legal voters can retrieve the voting credential. Though an adversary, say Bob, can record the encrypted messages by eavesdropping, we will show why this can not be used to construct a receipt later.



\begin{table}[t]
    \renewcommand{\arraystretch}{1.3}
    \caption{The Ideas in Our Proposed Protocol}

    \label{table:ideas}\small  

    \centering 

    \begin{tabular}[b]{p{3.4cm} p{4.7cm}}

        \toprule

        Goal &Method\\
        \midrule

        Authentication & Anonymous Credential \\

        Anonymity &Pseudonym\\

        Receipt-freeness &Key-private Proxy Re-encryption\\

        Efficiency $\&$ Mobility &Efficient Proxy Re-encryption\\

        \bottomrule

    \end{tabular}

\end{table}

Then, we account for how receipt-freeness is achieved. To achieve receipt-freeness, some voter-unknown randomness should be added into the ballot by voting authorities. A straightforward idea is that, by taking advantage of PRE scheme, ballot can be treated as re-encryptable ciphertext (a.k.a. second-level ciphertext) and cast to the bulletin board. Since the natural property of PRE, voting authority (proxy) can transform it to another one. However, if the re-encryption keys are stored on proxy server in advance, and Alice only cast the ciphertext anonymously, proxy will not know which re-encryption key to use. To make a repair, Alice need to generate the re-encryption key himself and cast it together with ballot to proxy. Unfortunately, two main defects rise with this repair. The main defect is that, provided both ciphertext and re-encryption key, any one including the adversary Bob can finish the transformation. Alice can, therefore, construct a receipt to Bob. Although it makes sense to allow Alice to vote directly and secretly to authority. Considered the eavesdropping problem, an untappable channel is inevitable. Furthermore, as another defect, one may doubt that voter-side computation cost is rather considerable.

Based on this, we present a more deliberate approach. In our protocol, ballot is treated as a re-encryption key, other than any ciphertext. This is possible since the generation of re-encryption key only involves the public key of preferred voting candidate and the short-term private key. Then Alice cast the ballot via an anonymous channel to the bulletin board. A hash value of the anonymous credential is as well enclosed for authentication. Proxy checks validity of the hash value and useability of the corresponding credential, and then encrypts an intelligible message with the pseudonym public key of Alice. Next proxy transforms it with the re-encryption key which Alice cast. The re-encrypted ciphertext will be published to bulletin board and can only be decrypted by the preferred candidate of Alice.




Stated informally, receipt-freeness is achieved from the following facts:
\begin{itemize}
  \item[a.] Proxy and adversaries cannot extract identities of voter and candidate from the re-encryption key (i.e. ballot).
  \item[b.] The voter does not know what message is encrypted and transferred to the voting transaction on bulletin board.
  \item[c.] Given the encryption form of credential, the ballot and the re-encrypted message to adversary, a voter cannot convince him in which way she voted.
  \item[d.] If adversaries coerce voters to disclose voting credential, voters can defraud coercers by presenting an indistinguishable fake.

\end{itemize}

\begin{figure*}[t]
    \begin{subfigure}{0.48\textwidth}
        \includegraphics[width=3.7in]{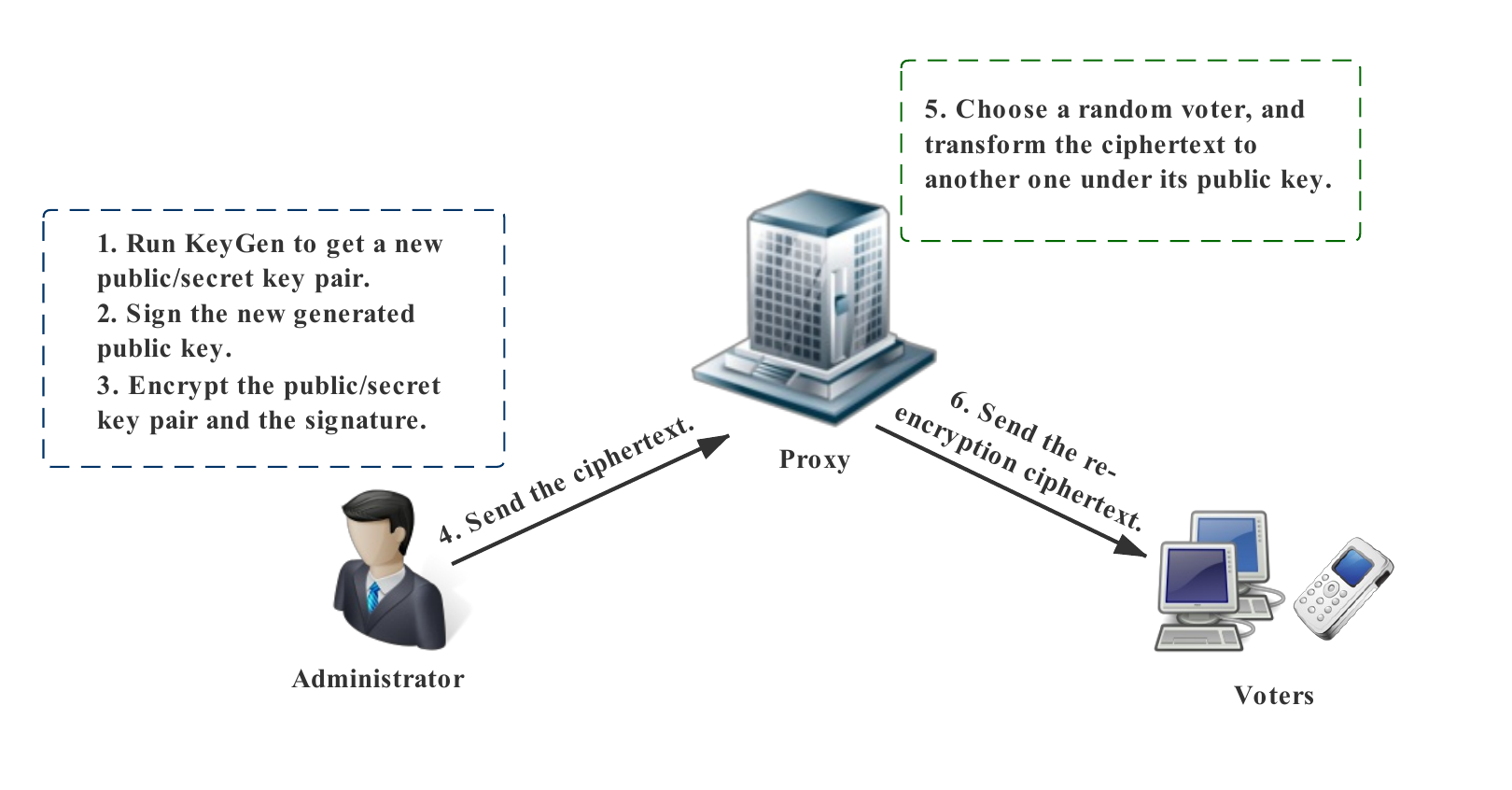}
        \caption{Credential Dispatching Stage.}
        \label{fig_Credential_Dispatching}
    \end{subfigure}
    \hspace*{\fill}
    \begin{subfigure}{0.48\textwidth}
        \includegraphics[width=3.7in]{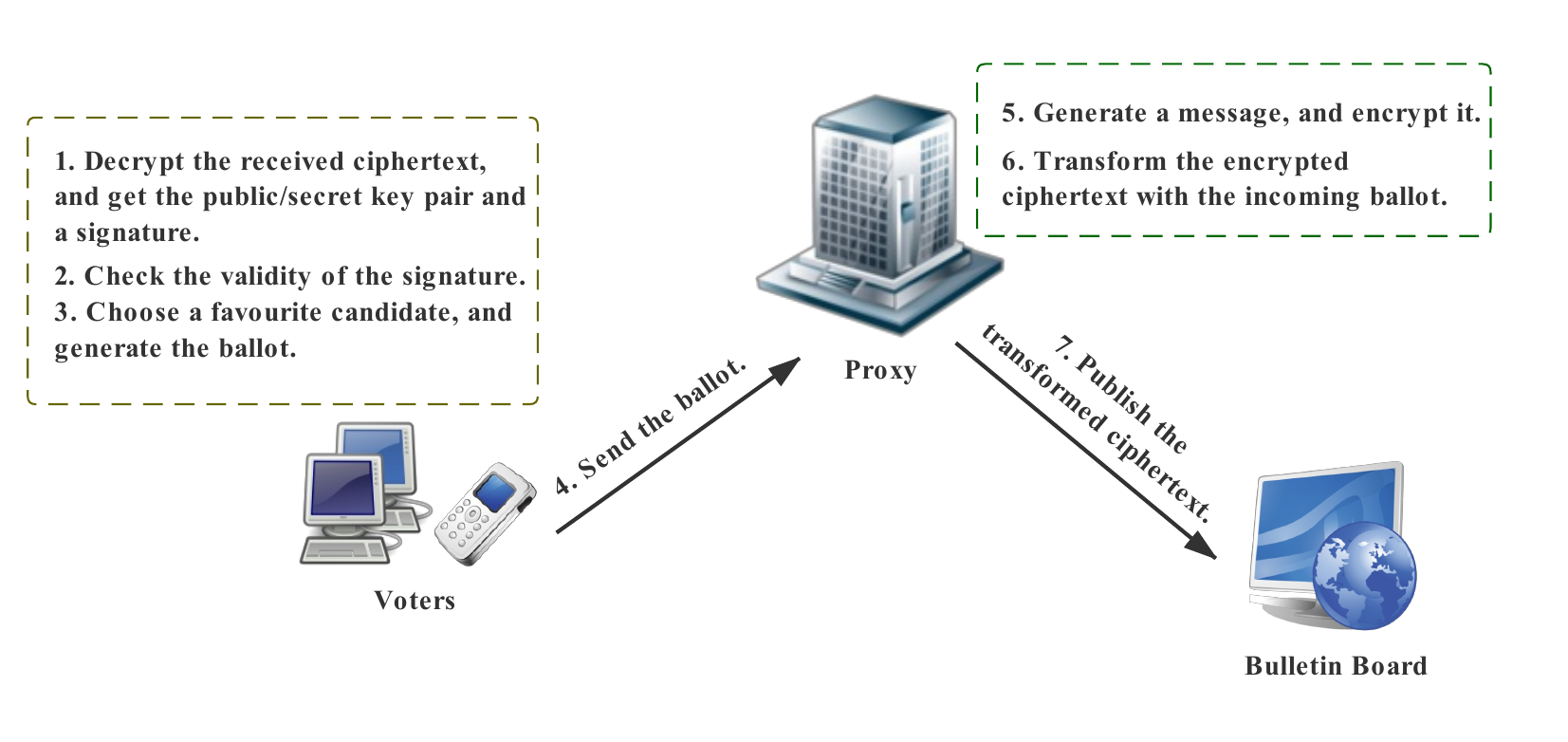}
        \caption{Ballot Casting Stage.}
        \label{fig_Ballot_Casting}
    \end{subfigure}
    \caption{High level description: How our voting protocol benefits from proxy re-encryption.} \label{fig_2}
\end{figure*}

To conclude, the main design goals with corresponding methods are shown in \Cref{table:ideas}. Moreover, to make our high level idea more comprehensive, some details are shown in \Cref{fig_Credential_Dispatching} and \Cref{fig_Ballot_Casting}.

\subsection{Notations}

For ease of description, some intuitive notations and abbreviations used in our proposed protocol are shown in \Cref{table:notations}.

\begin{table*}[t]
    \renewcommand{\arraystretch}{1.3}

    \caption{Notations used in the proposed voting scheme}

    \label{table:notations}\small  

    \centering 

    \begin{tabular}[b]{*{2}{p{3.8cm} p{13.5cm}<{\raggedright}}}

        \toprule

        Symbol & Description \\
        \midrule


         $\mathsf{par}$  & Public parameters of proxy re-encryption scheme and multi-designated verifiers signature scheme\\

         $\vec{L}_c$  & A list of candidates \\

         $\vec{L}_v$  & A list of legal voters \\

         $\vec{L}_k$  & A list of re-encryption keys from administrator to legitimate voters \\

         $\vec{L}_\sigma$       & A list of legitimate credentials \\

         $(y_E, x_E)$ & Long-term public/secret key pair of entity $E$ \\

         $(PK_i, SK_i)$  & Short-term public/secret key pair of index $i$ \\

         $\sigma$  & Voting credential \\

         $m$  & Voting message \\

         $\delta$  & Voting transaction \\

         $h_i$   &   The hash value of the $i^{th}$ voting credential \\

         $rk_{i \to j}$  & Re-encryption key from entity $i$ to entity $j$ \\

         $V_i$ & A voter, a.k.a. Alice \\

         $C_j$  & A candidate, a.k.a Bob \\

        \bottomrule

    \end{tabular}

\end{table*}

%
%
%
%
%
%
%
%
%
%
%
%
%
%
%
%
%
%


\subsection{Technical Details}
\subsubsection{System Setup}

\begin{description}
  \item[Step S1.] The administrator holds a database with $|\vec{L}_v|$ records of legitimate voter and publishes the following parameters to the bulletin board to set up the system:
\begin{itemize}
  \item[1)] The parameters of the re-encryption scheme and multi-designated verifiers signature scheme, i.e. $\mathsf{par}$.
  \item[2)] Self-signed public key certificate of administrator used in this round of voting. We denote such public key as $y_A$.
  \item[3)] A vector $\vec{L}_k$ of re-encryption keys from administrator to each voter.
  \item[4)] A kind of collision-free one-way hash function denoted as $Hash(\cdot)$.
\end{itemize}
  \item[Step S2.] Every candidate in the candidate slate $\vec{L}_c$ publishes its public key $y_{C_j}$ to the bulletin board.
\end{description}

\subsubsection{Credential Dispatching}

  For $i = 1, \dots, |\vec{L}_v|$, administrator executes:
    \begin{description}
      \item[Step C1.] Administrator generates a new pair of short-term public/secret keys $(PK_i, SK_i)$.
      \item[Step C2.] Administrator generates a multi-designated verifiers signature $s_i \gets Sign_{x_A, \vec{L}_v}(PK_i)$ using the described algorithm. This signature, together with the signed message $PK_i$, is the so-called \emph{credential}, which is denoted as $\sigma_i  = (PK_i, s_i)$.
      \item[Step C3.] Administrator encrypts the short-term secret keys $SK_i$ and credential $\sigma_i$ using the first-level Enc algorithm of PRE scheme $c_i \gets Enc_{y_A}(SK_i, \sigma_i)$.
      \item[Step C4.] Administrator sends the first-level encryption ciphertext $c_i$ to the proxy.
    \end{description}

%
%
%

  After receiving every ciphertext $c_i$ from administrator, the proxy executes the following steps:
    \begin{description}
      \item[Step C5.] Proxy randomly chooses an \emph{unused} number $j$ from $1$ to $|\vec{L}_v|$, and finds the corresponding re-encryption key $rk_{A \to V_j}$ from vector $\vec{L}_k$.
      \item[Step C6.] Proxy transfers ciphertext $c_i$ encrypted under administrator's public key $y_A$ to ciphertext under public key of the $l^{th}$ voter by using re-encryption algorithm $c_i^\prime \gets ReEnc_{rk_{A \to V_l}}(c_i)$.
      \item[Step C7.] Proxy sends the re-encrypted ciphertext $c_i^\prime$ to the $j^{th}$ voter.
    \end{description}

%
%

\begin{figure}[t]
    \centering
    \includegraphics[width=3.0in]{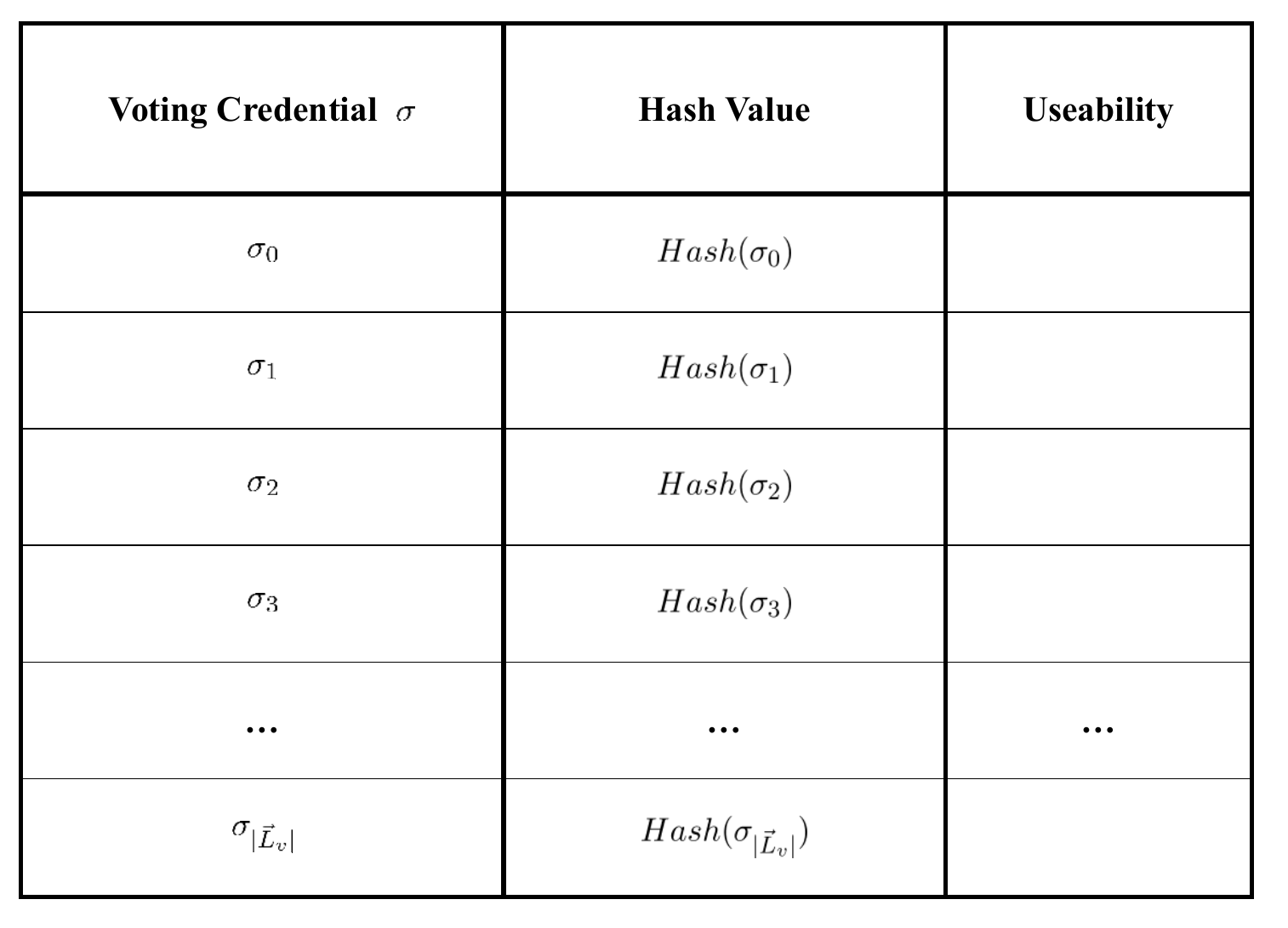}
    \caption{The list that proxy maintains.}
    \label{fig_list}
\end{figure}

  After all credentials are dispatched, administrator packs a list of voting credential and encrypts it with the public key of proxy as $Enc_{y_{P}}(\vec{L}_\sigma)$ and sends to the proxy. Proxy then decrypts it and maintain a hashtable (as shown in \Cref{fig_list}) of all voting credentials.



\subsubsection{Ballot Casting}

After receiving $c_i^\prime$, the $l^{th}$ voter executes the following steps:

    \begin{description}
      \item[Step B1.] Voter decrypts $c_i^\prime$ with its secret key $x_{V_l}$ and gets plaintext $(SK_i, \sigma_i) \gets Dec_{x_{V_l}}(c_i^\prime)$.
      \item[Step B2.] Voter checks the validity of the credential $\sigma_i$ by using multi-designated verifiers signature verification algorithm $Verify_{y_A}(PK_i, s_i)$. If the algorithm outputs 0, voter will abort voting process and may denounce the misbehavior of authorities.
      \item[Step B3.] Voter chooses a desired candidate $C_x$, and generates a ballot by using re-encryption key generation algorithm $Ballot \gets ReKeyGen(SK_i, y_{C_x})$.
      \item[Step B4.] Voter uses the given hash function to compute the hash value of the voting credential $h_i \gets Hash(\sigma_i)$.
      \item[Step B5.] Voter sends message $m_l = (Ballot, h_i)$ to the bulletin board via an anonymous communication channel.
    \end{description}

%
%
%
%

For every message $m_l$ which voter cast to the bulletin board, the proxy executes the following steps:

    \begin{description}
      \item[Step B6.] Proxy queries the maintained list of credentials. If hash value of the incoming credential is illegal or used, proxy will abort the process.
      \item[Step B7.] Otherwise, proxy uses current timestamp as plaintext $m_l^\prime \gets Stp$, and encrypts the chosen plaintext under public key $PK_i$ using the first-level Enc algorithm of PRE scheme $\delta_l \gets Enc_{PK_i}(m_l^\prime)$.
      \item[Step B8.] Proxy transfers ciphertext encrypted under public key $PK_i$ to ciphertext under public key of candidate $y_{C_x}$ by using re-encryption algorithm $\delta_l^\prime \gets ReEnc_{Ballot}(\delta_l)$. $\delta_l^\prime$ is the so called voting transaction.
      \item[Step B9.] Proxy publishes the voting transaction $\delta_l^\prime$ to bulletin board.
    \end{description}


%
%
%


\subsubsection{Ballot Opening and Tallying}
A candidate, say $C_j$, keeps continuous attention on voting transactions from bulletin board. When a new transaction, say $\delta_x$, is published, candidate $C_j$ attempts to decrypt it using its own secret key $x_{C_j}$. Once decryption succeeds, and the outcome is a timestamp, a ballot towards candidate $C_j$ is called to be opened. Candidate $C_j$ can keep secret that how many ballots it has received until date of tallying. At that date, administrator sends a message to bulletin board that separate the messages from different phases and inform candidates to publish their private key to bulletin board with their encoded identity like $\{ x_{C_j}, C_j \}$. After that, candidates publish as many valid messages as possible to bulletin board.

A tallying script on bulletin board will also be triggered to count the number of ballots of each candidate. After all ballots are tallied, messages that show "who received how many ballots" will also be published to bulletin board.

\subsubsection{Ballot Verifying and Auditing}
An additional auditing can optionally be implemented by election holder and any skeptical voter. This is easy to achieve because the set of private keys of candidates has been published on the bulletin board and all re-encrypted ballots can also be found.

\section{System Evaluation}
\label{sec_analysis}

\subsection{Analysis of Security Properties}
In this subsection, we give a heuristic analysis, as vast majority of previous voting protocols, on the security properties of our proposed scheme.

\textbf{Ballot secrecy.} Since the ballot in our proposed protocol is the re-encryption key in proxy re-encryption scheme, the secrecy of the ballot derives from the secrecy of re-encryption key. The secrecy of re-encryption key requires that adversaries (proxy) can not identify either delegator (voter) or delegatee (candidate) when given the re-encryption key $rk_{V \to C}$. This is achieved by the \emph{key-privacy} property in proxy re-encryption schemes. Additionally, since the anonymous credential that voter receive is different in each round of voting, ballot will not be the same even when the voter vote for the same candidate. Hence no external observation can obtain the relationship between a particular voter and vote.

\textbf{Unreusability and Eligibility.} Only if an unused valid credential is showed, the ballot would be accepted by proxy. If a credential is repeated, proxy will reject corresponding ballots. This design makes our proposal satisfy \emph{unreusability}. Since credentials are only dispatched to legal voters, and since the secure one-way hash function is collision-free, illegal voters can not off the hash value of a valid credentials which they are unknown. Accordingly, the illegal ones will easily be detected when casting their ballots. Hence \emph{eligibility} is achieved.

\textbf{Voter-fairness and Candidate-adaptiveness.} 
 Since the voting transactions are re-encrypted randomness under public key of candidate, no voter can learn about the voting outcome before the tallying phase. Hence \emph{voter-fairness} is achieved. Each candidate can learn how many votes are towards itself in the voting process. Candidate is able to have a direct understanding about what percentage his votes account for. However, he cannot explicitly know how many votes other candidates get. That means \emph{candidate-adaptiveness}.

\subsection{Analysis of Receipt-freeness and Coercion-resistance}
\textbf{Receipt-freeness Analysis:} \footnote{In the context of receipt-freeness, Alice cannot reveal her voting key, i.e. credential in our protocol.} In analysis receipt-freeness analysis, we first consider whether voter can construct a provable receipt by using materials which can be easily obtained by adversaries. Since all communication is carried via public channel, adversaries can readily obtain ballot $rk_{i \to C_j}$ from anonymous source, hash value of credential $Hash(\sigma_i)$, ciphertext under candidate's public key $Enc_{y_C}(\#)$ \footnote{$\#$ is just a placeholder.} and encryption form of credentials $Enc_{y_V}(\sigma_i)$. During the voting procedure, adversaries may try the following strategies:

\textbf{Strategy 1:} Adversary $\mathcal{A}$ may try to find out how Alice vote from what she cast to bulletin board, i.e. $rk_{i \to C_j}$ and $Hash(\sigma_i)$.

Only if $\mathcal{A}$ can retrieve $\sigma_i$, this attack will succeed. That is because, in many scenarios $\mathcal{A}$ is played by candidate, say $\widetilde{Bob}$, himself. Just as what proxy does, $\mathcal{A}$ can choose a particular string, encrypt it with the short-term public key in credential $\sigma_i$ and then transform it with the ballot $rk_{i \to C_j}$. Since $\mathcal{A}$  holds the private key of $\widetilde{Bob}$, he can easily verify whether the ballot is voting for $\widetilde{Bob}$. However, what $\mathcal{A}$ can obtain is $Hash(\sigma_i)$. Since the one-wayness property of Hash function, this strategy is not feasible in protocol.

\textbf{Strategy 2:} Adversary $\mathcal{A}$ may try to find out how Alice vote from her ballot and the processed voting transaction, i.e. $rk_{i \to C_j}$ and $Enc_{y_C}(\#)$.

$\widetilde{Alice}$, the tamed voter, involves in this attack. $\mathcal{A}$ bribes $\widetilde{Alice}$ for her support and requires her proof. However, as described in \Cref{sec_proposal} and \ref{APP_PRE_Details}, the final ciphertext published on the bulletin board is in form of $Enc_{y_C}(\#) = \delta_j^\prime = ReEnc_{Ballot}(\#) = (t_1^\prime, t_2^\prime) = (Z^{a_{j2}y}, \# \cdot Z^{y})$, where $y = k(a_{i1} + r) + ww^\prime$ (\Cref{equ_rekeygen}). Since the random factors $k, \#, w^\prime \in_R \mathbb{Z}_q$ is brought in by proxy, $\widetilde{Alice}$ has no knowledge about that which message on bulletin board is correlated to her own choice. Thus, $\widetilde{Alice}$ cannot convince $\mathcal{A}$ that she voted as he wishes.


\textbf{Coercion-resistance Analysis:} $Laoco\ddot{o}n$ also achieves a stronger notion of receipt-freeness which is entitled with coercion-resistance in Juels et al.'s protocol \citep{juels2005coercion}. In the model of coercion-resistance, the capability of adversary $\mathcal{A}$ has been extremely amplified. $\mathcal{A}$ can coerce Alice to divulge her voting credential and coerce proxy to leak what randomness it brought in during the process. To sum up, Jeuls et al. advance three types of attack that coercion-resist protocol should defend:

\textbf{Randomization Attack:} Adversary $\mathcal{A}$ can coerce Alice by requiring that she submit randomly composed balloting material. This attack aims to nullify voting result with a large probability and works in precincts where competitors gains more popularity.

\textbf{Forced-abstention Attack:} Adversary $\mathcal{A}$ can coerce Alice by demanding that she refrain from voting. This attack works when authentication is in a direct and public approach.

\textbf{Simulation attack:} Adversary $\mathcal{A}$ can coerce Alice by causing her to divulge her private keying material after the registration process but prior to the election process. Then $\mathcal{A}$ can cast the ballot in name of Alice.

\textbf{Our defence:} $Laoco\ddot{o}n$ defend the above attacks by allowing voters to generate fake credentials which is indistinguishable to possible adversaries. Since the integrity of credentials is promised by designated verifiers signature, adversary who is outside the set of verifiers cannot distinguish whether the coerced credential is fake or not. If coercion happens multiple times, voter can simply release the same fake credential $\widetilde{\sigma}$. Thus, our proposed protocol is secure against coercion attacks.

\begin{table*}[hbt]
    \renewcommand{\arraystretch}{1.3}
    \caption{Analysis of Efficiency}

    \label{table:efficiency}\small  

    \centering 

    \begin{threeparttable}
%
    \begin{tabular}[b]{*{6}{p{4.0cm} p{2.1cm}<{\raggedright}}}

        \toprule 
        Phases                  & Entities  & Computation Cost & Time Spent & Communication Cost  \\
        \midrule
        \multirow{2}{*}{Credential Dispatching}   &    Administrator  &    $3E_1 + 2E_2 + S$ + Sig  &   20.4ms + $time_{S}$ &  $2|\mathbb{G}_1| + |\mathbb{G}_T|$   \\
                                                  \cmidrule{2-2}  \cmidrule{3-3} \cmidrule{4-4} \cmidrule{5-5}
                                                  &    Proxy          &  $2E_2 + 3S + 4P$            &   24.9ms &  $2|\mathbb{G}_T|$\\
        \multirow{2}{*}{Ballot Casting}           &    Voter     &    $E_1 + 4E_2 + P$ + Vfy         &   14.7ms + $time_{V}$ &  $3|\mathbb{G}_1| + 3|\mathbb{G}_T| + |Hash|$ \\
                                                  \cmidrule{2-2} \cmidrule{3-3} \cmidrule{4-4} \cmidrule{5-5}
                                                  &    Proxy     &   $2E_1 + 3E_2 + 4S + 4P$   &   38.3ms &  $2|\mathbb{G}_T|$\\
        \multirow{1}{*}{Ballot Opening}           &    Candidate     &    $E_2$   &   0.6ms   &  $|\mathbb{Z}_q|$\\

        \bottomrule
    \end{tabular}

    \begin{tablenotes}
    \item[*] Measurement: per vote.
    \end{tablenotes}

    \end{threeparttable}

\end{table*}

\begin{table*}[h]
    \renewcommand{\arraystretch}{1.3}
    \caption{Property Comparison of Different Receipt-freeness Voting Protocols}

    \label{table:func_comparison}\small  

    \centering 

    \begin{threeparttable}
    \begin{tabular}[b]{p{4.0cm} *{2}{p{1.7cm}}  *{4}{p{2.0cm}<{\raggedright}}}

        \toprule
       \multirow{2}{*}{Functionalities}    & Protocols\\\cmidrule{2-7}
                               & \cite{okamoto1997receipt} & \cite{lee2002receipt} &  \cite{juels2005coercion} & \cite{chow2008robust} & \cite{xia2018framework} & Our protocol \\
        \midrule
        Receipt-freeness         &    $\checkmark$     &    $\checkmark$ & \checkmark  &   $\checkmark$     &   $\checkmark$    & $\checkmark$ \\
        Unreusability &    $\checkmark$     &    $\checkmark$   & \checkmark &  $\checkmark$     &   $\checkmark$    & $\checkmark$ \\
        Ballot Secrecy           &    $\checkmark$     &    $\checkmark$ & \checkmark  &   $\checkmark$     &   $\checkmark$    & $\checkmark$ \\
        Vote-and-go              &    $\times$     &    $\times$   & \checkmark  & $\checkmark$     &   $\checkmark$    & $\checkmark$ \\
        Coercion-resistance      &   $\times$  &   $\times$   & $\checkmark$  &   $\times$   &  $\times$ & $\checkmark$ \\
        Efficiency               &    $\checkmark$     &    $\checkmark$   & \checkmark & $\checkmark$     &   $\times$    & $\checkmark$ \\
        Scalability              &    $\checkmark$     &    $\checkmark$   & \checkmark & $\times$     &   $\times$    & $\checkmark$ \\
        Mobility                 &    $\times$     &    $\times$   &   $\checkmark$  & \checkmark   &   $\times$    & $\checkmark$ \\
        Light-weight \tnote{a}   &    $\times$     & $\times$  &   $\times$ &   $\checkmark$  & $\checkmark$  &      $\checkmark$ \\
        Verifiability            &    $\times$     & $\checkmark$  &   $\times$ &   $\checkmark$  & $\checkmark$  &$\bigcirc$ \tnote{b}\\
        Fairness                 &    $\checkmark$     &    $\checkmark$   & \checkmark & $\checkmark$     &   $\checkmark$    & $\times$ \\
        Voter-fairness           &    $\times$     &    $\times$   &   $\times$  &   $\times$   &   $\times$    & $\checkmark$ \\
        Candidate-adaptiveness   &    $\times$     &    $\times$   &   $\times$  &   $\times$   &   $\times$    & $\checkmark$ \\
        Without Untappable Channel    & $\times$ & \checkmark \tnote{c} &   $\checkmark$  & $\times$  & $\times$ &  $\checkmark$ \\

        \bottomrule
    \end{tabular}
    \begin{tablenotes}
        \item[a] It refers to as light-weight voter-side calculation.
        \item[b] Note that $\bigcirc$ denotes extensible.
        \item[c] Their protocol takes advantage of trusted hardware.
    \end{tablenotes}

    \end{threeparttable}

\end{table*}

\subsection{Analysis of Efficiency and Comparison}

In this subsection, we analysis the efficiency of our proposed voting protocol in terms of computation cost and communication cost. The result can be seen in Table \ref{table:efficiency}. In the analysis, we only take time spent on pairings and exponentiations into consideration, since the complexity of hash function, multiplications or additions in finite cyclic groups is relatively negligible. We denote by $E_1$ a exponentiation in $\mathbb{G}$, by $E_2$ a exponentiation in $\mathbb{G}_T$, by $P$ a pairing operation. Sig and Vfy respectively denote the signature and verification in digital signature scheme. For ease of comprehension, we also give the concrete time spent in \Cref{table:efficiency}, whose data are from references \citep{shao2012anonymous,shi2007multi,kiltz2007chosen}. Accordingly, time spent on one exponentiation in $\mathbb{G}$, one exponentiation in $\mathbb{G}_T$ and one pairing are respectively 6.4, 0.6 and 5.9 ms. These benchmarks are measured on a workstation whose processor is  a 64-bit, 3.2 GHz Pentium 4. Note that groups $\mathbb{G}, \mathbb{G}_T$ all have a order of 160-bit where the former two groups are used in proxy re-encryption scheme. One element in $\mathbb{G}, \mathbb{G}_T$ is 512-bit length. Since the construction of MDVS scheme is not related to the security of our protocol, we will not delve deep into its algorithm details and denote the time cost of signature and verification respectively as $time_{S}$ and $time_{V}$.

We also consider the communication cost of our proposed protocol. As shown in \Cref{table:efficiency}, our protocol enjoys a rather small communication cost. Since only messages which are published by proxy in \textbf{Ballot Casting} phase and by candidate in \textbf{Ballot Opening} phase will be recorded on the bulletin board, the size of the whole bulletin board is linear in the amount of voters $|\vec{L}_v|$. That is, the complexity of tallying, $\mathcal{O}(|\vec{L}_v|)$, is linear to voter account. After the overall consideration of computation and communication cost, our proposed e-voting protocol is suitable for large scale elections (a.k.a. referendums). And since the low overhead of voter side computation, and the no restriction on the designated location in which voter casts its ballot, our protocol also achieves \emph{mobility}. That means our protocol can be implemented into a vote-by-mobile system.

\label{sec_comp}
We also compare the properties in our proposed protocol with some representative protocols reviewed in \Cref{reviews}. We give out our comparison in \Cref{table:func_comparison}.

\subsection{Towards Additional Properties}

\textbf{Individual Verifiability.} Individual verifiability refers to the capability of individual voters to confirm their choice has been correctly encoded in voting script and counted in tallying result \citep{ryan2009pret}. Note that, \emph{verifiability} is somewhat contradicting to \emph{receipt-freeness} \citep{wang2017review}. We proposed a non-verifiability version above in \Cref{sec_proposal}. Here we find a middle ground between verifiability and receipt-freeness and show how individual verifiability can be extended to achieve in our proposed protocol.

This extension takes advantage of bit commitment scheme \citep{naor1991bit} used in Fujioka et al.'s voting protocol \citep{fujioka1992practical}. In \textbf{Ballot Casting} phase, proxy makes a bit commitment about the voting credential using a randomly chosen key $k$. The process is denoted as $\beta_i \gets \xi_k(\sigma_i)$. In a round of election, the key used for bit commitment is the same to all different ballot. Enclosed with the voting transaction, the commitment remains secrecy until \textbf{Ballot Verifying and Auditing} phase. Proxy publishes the key $k$ to bulletin board in this phase and all voter can confirm whether its ballot is correctly encoded and counted in the voting process. Additionally, for a more user-friendly approach, one may consult \citep{ryan2016selene,iovino2017using}

\section{Conclusion and Future Work}
\label{sec_conclusion}
In this paper, we introduced key-private proxy re-encryption as a new cryptographic tool to construct e-voting protocols. Based on this special PRE scheme, we proposed $Laoco\ddot{o}n$, a practical and receipt-free protocol for large scale remote electronic voting. Additionally, in our protocol, candidates knows their approval rating throughout the voting process. Due to receipt-freeness, candidates are allowed to rise legal voting campaigns except for any bribing or coercion. Experimental results show that our protocol enjoys a light-weighted voter-side calculation and is efficient enough to deploy in a referendum.

Finally, we acknowledge that there are still many possible and attractive problems that need further investigation.
\begin{itemize}
  \item[1.] $Laoco\ddot{o}n$ has one deficiency, where the voting authorities are centralized. It will be fragile to defend a (distribute) denial-of-service (DoS/DDoS) attack and not robust when adversary corrupts all voting authorities. Many voting protocols address these attacks in a threshold manner. It remains an interesting problem that whether $Laoco\ddot{o}n$ can be extended to a multi-authorities version. As a simple envision, multi-use or threshold PRE schemes can be utilized.

  \item[2.] Few study in proxy re-encryption has considered the potential of PRE to construct a voting system. The PRE scheme used in our work is just CPA secure and based on bilinear pairings. Our voting protocol can be more efficient and secure as the result of using better PRE schemes. Hence, identifying which particular kind of PRE is suitable for voting and proposing more efficient and secure PRE schemes of that kind can be regarded as another open problem.
      
  \item[3.] Some of the existing e-voting protocols \citep{unruh2010universally,alwen2015incoercible} provided a more rigorous proof of the security guarantees. These protocols are adapted in the universal composability (UC) framework \citep{canetti2001universally}. Evaluating security of our e-voting protocol in the UC framework is also considered as our future work.

\end{itemize}



%

\appendix
\section{Algorithm Details of a Key-private Proxy Re-encryption Scheme \citep{ateniese2009key}}
\label{APP_PRE_Details}
An efficient single-use unidirectional key-private PRE scheme \textbf{SU-KP-PRE} = (\textbf{Setup, KeyGen, ReKeyGen, Enc, ReEnc, Dec}) consists of six algorithms. The detailed description is as follows.

\textbf{Setup:} $\mathsf{par} \gets \textbf{Setup}(1^k)$. On inputting the security parameter $k$, the outputting system parameters are $(g, h, q, \mathbb{G}, \mathbb{G}_T, \hat{e}, Z)$, where $\mathbb{G}$ is a finite cyclic group generated by $g$. And $h$ is another random generator of $\mathbb{G}$. $\mathbb{G}$ and $\mathbb{G}_T$ are of prime order $q$. And $\hat{e}$ is a efficient bilinear map such that $\hat{e}$: $\mathbb{G} \times \mathbb{G} \to \mathbb{G}_T$. $Z$ is computed by $Z = \hat{e}(g, h)$.

\textbf{KeyGen:} $(PK_i, SK_i) \gets \textbf{KeyGen}(\mathsf{par})$. The public key is set as $PK_i = (pk_{i1}, pk_{i2}) = (Z^{a_{i1}}, g^{a_{i2}})$ for random $a_{i1}, a_{i2} \in_R \mathbb{Z}_q$. Here $(a_{i1}, a_{i2})$ is the corresponding secret key.

\textbf{ReKeyGen:} $rk_{i \to j} \gets \textbf{ReKeyGen}(SK_i, PK_j)$. On inputting a public key $PK_j = (pk_{j1}, pk_{j1}) = (Z^{a_{j1}}, g^{a_{j2}})$ and a secret key $SK_i = (a_{i1}, a_{i2})$, the re-encryption key $rk_{i \to j}$ is constructed as follows.
    \begin{enumerate}
      \item Randomly choose $r, w \in_R \mathbb{Z}_q$.

      \item Compute

      \begin{equation}\label{equ_rekeygen}
      \begin{split}
        rk_{i \to j}  & = ((g^{a_{j2}})^{a_{i1}+r}, h^r, \hat{e}(g^{a_{j2}}, h)^w, \hat{e}(g, h)^w)\\
        & = ((g^{a_{j2}})^{a_{i1}+r}, h^r, Z^{a_{j2}w}, Z^w)
      \end{split}
      \end{equation}

    \end{enumerate}

\textbf{Enc:} $c_i \gets \textbf{Enc}_{PK_i}(m)$. On inputting a public key $PK_i = (pk_{i1}, pk_{i2}) = (Z^{a_{i1}}, g^{a_{i2}})$ and a message $m \in \mathbb{G}_T$, the following steps are calculated by encryptor:
    \begin{enumerate}
      \item Randomly choose $k \in_R \mathbb{Z}_q$.

      \item Compute the ciphertext $c_i = (g^k, h^k, m \cdot Z^{a_{i1}k})$.
    \end{enumerate}

\textbf{ReEnc:} $c_j \gets \textbf{ReEnc}_{rk_{i \to j}}(c_i)$. On inputting a re-encryption key $rk_{i \to j} = (R_1, R_2, R_3, R_4) = ((g^{a_{j2}})^{a_{i1}+r}, h^r, Z^{a_{j2}w}, Z^w)$ and a ciphertext $c_i = (\alpha, \beta, \gamma) = (g^k, h^k, m \cdot Z^{a_{i1}k})$ under public key $PK_i$, proxy transforms the second-level ciphertext into a first-level one as follows.
    \begin{enumerate}
      \item Check whether $\hat{e}(\alpha, h) = \hat{e}(g, \beta)$. If the equation does not hold, the algorithm outputs $\perp$ and aborts; otherwise there exists some $k \in \mathbb{Z}_q$ and $m \in \mathbb{G}_T$ such that $\alpha = g^k$, $\beta = h^k$ and $ \gamma = m \cdot Z^{a_{i1}k}$, and thus, $c_i = (\alpha, \beta, \gamma)$ is a valid encryption of message $m$ under $PK_i  = (Z^{a_{i1}}, g^{a_{i2}})$

      \item Compute $t_1  = \hat{e}(R_1, \beta)  = \hat{e}(g^{a_{j2}(a_{i1}+r)}, h^k)  = Z^{a_{j2}k(a_{i1}+r)}$.

      \item Compute $t_2= \gamma \cdot \hat{e}(\alpha, R_2)= m \cdot Z^{a_{i1}k} \cdot \hat{e}(g^k, h^r)= m \cdot Z^{a_{i1}k} \cdot Z^{rk}= m \cdot Z^{k(a_{i1}+r)}$.

      \item Randomly choose $w^\prime \in_R \mathbb{Z}_q$.

      \item Re-randomize $t_1$ by setting $t_1^\prime = t_1 \cdot R_3^{w^\prime} = Z^{a_{j2}k(a_{i1}+r)} \cdot (Z^{wa_{j2}})^{w^\prime} = Z^{a_{j2}(k(a_{i1}+r)+ww^\prime)}$.

      \item Re-randomize $t_2$ by setting $t_2^\prime = t_2 \cdot R_4^{w^\prime} = m \cdot Z^{k(a_{i1}+r)} \cdot (Z^{w})^{w^\prime} = m \cdot Z^{k(a_{i1}+r)+ww^\prime}$.

      \item Output the re-encrypted ciphertext $c_j = (t_1^\prime, t_2^\prime) = (Z^{a_{j2}y}, m \cdot Z^y)$, where $y = k(a_{i1}+r)+ww^\prime$.
    \end{enumerate}

\textbf{Dec:} $m \gets \textbf{Dec}_{SK_i}(c_i)$. On inputting secret key $SK_i = (a_{i1}, a_{i2})$ and any ciphertext $c_i$ under public key $PK_i$:
    \begin{itemize}
      \item[-] If $c_i$ is an original ciphertext (a.k.a. second-level ciphertext), such as $c_i = (\alpha, \beta, \gamma) = (g^k, h^k, m \cdot Z^{a_{i1}k})$, the decryption algorithm execute the following steps:
            \begin{enumerate}
              \item Check whether $\hat{e}(\alpha, h) = \hat{e}(g, \beta)$. If the equation does not hold, the algorithm outputs $\perp$ and aborts; otherwise goes to next steps.

              \item Output $m = \gamma / \hat{e}(\alpha, h)^{a_{i1}}$.

              \item Note that $m$ may be $\perp$.
            \end{enumerate}

      \item[-] If $c_i$ is an re-encrypted ciphertext (a.k.a. first-level ciphertext), such as $c_i = (\alpha, \beta) = (Z^{a_{i2}y}, m \cdot Z^y)$, output $m = \beta / \alpha^{1/a_{i2}}$ as the result.
    \end{itemize}


%
%
%
%

\section*{References}

\bibliography{ref}

\end{document}